\documentclass[manuscript]{aastex}

\slugcomment{}

	\shorttitle{}
	\shortauthors{}

\begin{document}


\title{Thermal Shadows and Compositional Structure in Comet Nuclei}


\author{Aur\'elie Guilbert-Lepoutre}
\affil{Department of Earth and Space Sciences, UCLA, Los Angeles, CA 90095}
\email{aguilbert@ucla.edu}

\and

\author{David Jewitt}
\affil{Department of Earth and Space Sciences, UCLA, Los Angeles, CA 90095}
\affil{Department of Physics and Astronomy, UCLA, Los Angeles, CA 90095}
\email{jewitt@ucla.edu}

\begin{abstract}
We use a fully 3-dimensional thermal evolution model to examine the effects of a non-uniform surface albedo on the subsurface thermal structure of comets.  Surface albedo markings cast ``thermal shadows'', with strong lateral thermal gradients. Corresponding compositional gradients can be strong, especially if the crystallization of amorphous water ice is triggered in the hottest regions. We show that the spatial extent of the structure depends mainly on the obliquity, thermal conductivity and heliocentric distance. In some circumstances, subsurface structure caused by the thermal shadows of surface features can be maintained for more than 10~Myr, the median transport time from the Kuiper Belt to the inner solar system. Non-uniform compositional structure can be an evolutionary product and does not necessarily imply that comets consist of building blocks accumulated in different regions of the protoplanetary disk.
\end{abstract}


\section{Introduction}
Solar system comets are currently stored in two main reservoirs, namely the Oort Cloud and the Kuiper Belt, having different dynamical histories and physical properties. The bodies contained in these reservoirs can be scattered to the inner solar system by various gravitational processes. 
The Scattered Disk component of the Kuiper Belt is widely believed to be the source of Jupiter Family Comets \citep[hereafter JFCs,][]{Vol08}. In particular, a gravitational cascade might exist between the three distinct populations: scattered disk objects, Centaurs and JFCs \citep{Lev97,Tis03}.

Comets are believed to contain some of the best-preserved material from the formation of our planetary system. Cometary composition should reflect the location at which the material --ices and dust-- formed in the protoplanetary disk. A gradient in composition might reasonably be expected, distinguishing comets formed at high temperatures close to the proto-Sun, from those formed in an environment dominated by interstellar chemistry at large heliocentric distances. This simple picture is complicated by potential radial mixing inside the protoplanetary disk. On a micro-scale, the co-existence of crystalline silicates (formed at temperatures $\geq$~10$^3$~K) with cometary ice (accreted at temperatures $\leq$~50~K) provides direct evidence for radial mixing \citep{Cam89, Ish08}. On a macro-scale, cometesimals formed at different heliocentric distances and temperatures might have been scattered and later aggregated to form individual cometary nuclei.
For this reason, one key question in cometary science is whether the comet nuclei observed today are internally homogeneous or heterogeneous in composition. Observationally, both types of chemical structure have been reported among the JFCs. For example outgassing from comets 9P/Tempel~1 \citep{Mum05, Fea07} and 103P/Hartley 2 (A'Hearn et al.~2011) appears to be compositionally non-uniform, whereas comet 73P/Schwassmann-Wachmann~3 is found to be uniform \citep{Del07, Kob07}.

However, several post-accretion processes could alter the primordial compositions of comets. Their surface layers might be chemically stratified by solar wind and cosmic ray irradiation during their 4.5~Gyr residence in the Kuiper Belt and the Oort Cloud reservoirs \citep{Coo03}. In addition, their internal structures and compositions could be locally modified by heat absorbed at the surface from the Sun (for example, images of the nucleus of comet 103P/Hartley~2 show local albedo variations up to a factor of 4 \citep{AHe11}.
The diversity in composition observed in comets \citep{AHe95} could consequently be the result of a combination of these three effects: different formation environments, chemical evolution during multi-billion year storage in the source reservoirs, and recent thermal processing by absorbed sunlight once injected into the planetary region.

In this paper, we use a new thermal evolution model that has been developed  to allow fast, accurate computation of the 3-dimensional heat transport problem \citep{Gui10}. The speed of the model allows us to accurately calculate the effects of non-uniform surface albedo on the internal structure of a cometary nucleus and to explore the influence of orbital and thermophysical parameters. The model and assumptions are presented in Section 2. The results for different configurations are presented in Section 3 and discussed in Section 4.

\section{Thermal evolution model}
\subsection{Main equations}
Ours is a "toy-model" in which the parameters of the nucleus are idealized in order to make the problem tractable, and no attempt is made to model any particular real nucleus. 
The body is assumed to be initially a sphere made of a porous mixture of ice and dust uniformly distributed within the icy matrix. \citet{Jew09} and \citet{Mee09} report observational, albeit indirect, evidence consistent with the presence of amorphous water ice in comets. We therefore assume the ice is initially amorphous. 
The model we use evaluates the temperature distribution inside the body by taking into account three dimensional heat fluxes, and includes energy release from the crystallization of amorphous ice. 
This 3-dimensional model is fully described in \citet{Gui10}. We here give only an outline. The heat conduction equation to be solved is the following:
\begin{equation}
\label{eq_depart}
\rho _{bulk}c~\frac{\partial T}{\partial t}~+~\nabla(- \kappa~ \overrightarrow{\nabla}~T)~ =~\mathcal{Q}_{cryst},
\end{equation}
where $T$ [K] is the temperature distribution to be determined, $\rho _{bulk}$ [kg m$^{-3}$] the object's bulk density, $c$ [J kg$^{-1}$ K$^{-1}$] the material heat capacity, $\kappa$ [W m$^{-1}$ K$^{-1}$] its effective thermal conductivity (parameters described next section), and $\mathcal{Q}_{cryst}$ [W m$^{-3}$] the internal power production per unit volume due to the amorphous-crystalline phase transition. The latter is described by:
\begin{equation}
\mathcal{Q}_{cryst} = \lambda (T)~ \rho _{a}~ H_{ac},
\end{equation}
with $\rho _a$ [kg m$^{-3}$] the amorphous water ice bulk density. The phase transition releases a latent heat $H_{ac}$ = 9$\times$10$^{4}$~J~kg$^{-1}$ \citep{Kli81}, at a rate measured by \citet{Sch89}:
\begin{equation}
\lambda (T) = 1.05 \times 10^{13} ~ e^{-5370/T} ~~s^{-1}.
\end{equation} 

Boundary conditions are considered both at the surface and in the center of the object. 
Several thermal processes are considered to evaluate the thermal balance for each point on the surface:
\begin{itemize}
\item[-] solar illumination described by $(1-\mathcal{A})\frac{S_{\odot}}{d^2_H}\cos\xi$, with $\mathcal{A}$ the Bond albedo, $S_{\odot}$ the solar constant, $d_H$ the object's heliocentric distance, and $\xi \leq 90^{\circ}$ the local zenith angle.
\item[-] thermal emission $\varepsilon \sigma T^4$, with $\varepsilon$ material emissivity, $\sigma$ the Stefan-Boltzmann constant and $T$ the surface temperature. 
\item[-] lateral and radial heat fluxes.
\end{itemize}

The heat diffusion equation (\ref{eq_depart}) is expanded in spherical coordinates:
\begin{equation}
\frac{\rho_{bulk} c}{\kappa}~\frac{\partial T}{\partial t} - \left( \frac{2}{r} + \frac{1}{\kappa}\frac{\partial \kappa}{\partial r}\right) \frac{\partial T}{\partial r} - \frac{\partial^2 T}{\partial r^2} -\frac{1}{r^2} \Delta_{\theta, \varphi}T = \frac{\mathcal{Q}_{cryst}}{\kappa},
\end{equation}
with $\Delta_{\theta, \varphi}$ the angular Laplacian operator. 
As spherical harmonics $\mathcal{Y}_{lm}$ allow a simple and natural expression of the temperature over a regular spherical grid, we introduce them to describe the temperature distribution:
\begin{equation}
\label{temp-approx}
T= \sum_{l=0}^{\infty} \sum_{m=-l}^l ~T^{lm}(t,r) ~\mathcal{Y}_{lm}(\theta , \varphi).
\end{equation}
This sum is exact as long as the degree $l$ goes to infinity. This cannot be reached in practise and the sum is thus cut to a maximum degree $l_{max}$. We then introduce the expansion in the heat diffusion equation:
\begin{equation}
\label{eq-finale}
\frac{\rho c}{\kappa} ~\frac{\partial T^{lm}}{\partial t} -\left(\frac{2}{r} +\frac{1}{\kappa}\frac{\partial \kappa}
{\partial r}\right)~\frac{\partial T^{lm}}{\partial r} - \frac{\partial^2 T^{lm}}{\partial r^2} + \frac{l(l+1)}{r^2}~T^{lm} = \mathcal{Q}_{cryst}^{lm},
\end{equation}
with $\mathcal{Q}_{cryst}^{lm}=\sqrt{4 \pi}~\mathcal{Q}_{cryst}~ \delta _{l,0}~ \delta _{m,0}$, $\delta$ being the Kronecker function. We therefore obtain $(l_{max} +1)^2$ equations of $T^{lm}(t,r)$, instead of one single 3-dimensional equation for $T$. These 1D equations are solved using a Crank-Nicholson numerical scheme, which is a stable implicit technique.

The boundary condition at the surface is given by the thermal balance evaluated for each point of the surface, expanded into the basis of spherical harmonics:
\begin{equation}
T^{lm}_{surf}= \int _0^{2\pi}\!\!\! \int _0^{\pi} T_{surf}(\theta , \varphi)~\mathcal{Y}_{lm}(\theta , \varphi) \sin \theta~ d\theta \, d \varphi.
\end{equation}
The sampling theorem developed by \citet{Dri94} is used to derive these coefficients. Denoting by $N$ the number of points in one direction of the equally sampled surface grid, the boundary conditions $T^{lm}_{surf}$ are computed as: 
\begin{equation}
T^{lm}_{surf}= \sum_{j=0}^{N-1} \sum_{k=0}^{N-1} ~a_j^{N/2}~T_{surf}(\theta _j, \varphi _k) ~\mathcal{Y}_{lm}(\theta _j , \varphi _k),
\end{equation} 
with $\theta _j = \frac{j \pi}{N}$, $\varphi _k = \frac{2k \pi}{N}$ the grid point coordinates, $a_j^{N/2}$ a coefficient that accounts for the over-sampling near the poles, and $T_{surf}$ [K] the equilibrium temperature at each point of the surface. 
The number of points $N$ in one angular direction is chosen so as to minimize the discretization errors and the computational load.
In the center, the boundary condition is simpler and depends only on $r$:
\begin{eqnarray}
\frac{\partial T(t,r,\theta,\varphi)}{\partial r}=0~\Longrightarrow~\frac{\partial T^{lm}(t,r)}{\partial r}=0.
\end{eqnarray}

\subsection{Thermophysical properties}
The bulk density is related to the porosity of the solid matrix $\psi$ by:
\begin{equation}
\rho _{bulk} = (1 - \psi) \left( \frac{X_{H_2O}}{\rho _{H_2O}}+\frac{X_{d}}{\rho _{d}} \right) ^{-1},
\end{equation}
with $X_{H_2O}$ and $X_d$ the mass fractions of water ice and dust respectively, $\rho_{H_2O}$ and $\rho_d$ [kg m$^{-3}$] the densities of water ice and dust respectively. 
We assume that the object is made of a material with a dust to water ice mass ratio $X_{d}/X_{H_2O}$ = 1, a porosity $\psi$ = 30\%, and a bulk density $\rho _{bulk}$ = 1~g~cm$^{-3}$. 
The heat capacity of the mixture is obtained by computing the average of the values weighted by the mass fraction of each component
\begin{equation}
c=X_{H_2O} c_{H_2O} + X_d c_d,
\end{equation}
with X$_{H_2O}$ and X$_d$ the mass fraction of water ice and dust, and c$_{H_2O}$ and c$_d$ [J kg$^{-1}$ K$^{-1}$] the heat capacities of each component. The numerical values used in this work can be found in Table \ref{param}.

We evaluate the thermal conductivity by considering the material as made of two phases, the empty pores with a thermal conductivity $\kappa _{p}$ and the solid matrix with a thermal conductivity $\kappa _{s}$. Within the empty pores the heat is transferred through thermal radiation for which the effective conductivity is
\begin{equation}
\kappa _{p}=4 r_{p} \varepsilon \sigma T^{3},
\end{equation}
with $r_p$ =1~$\mu$m the average pore radius, $\varepsilon$=0.9 the medium emissivity, $\sigma$ the Stefan-Boltzmann constant and $T$ [K] the temperature \citep{Hue06}.
The solid matrix thermal conductivity, $\kappa_s$, is computed as the average of each component thermal conductivity (see Table \ref{param}), weighted by its volume fraction:
\begin{equation}
\kappa _{s} = x_{H_2O} \left[(1-X_{cr}) \kappa _{a}+X_{cr} \kappa _{cr} \right]+x_{d} \kappa _{d},
\end{equation}
with $x_{H_2O}$ and $x_d$ the volume fractions of water ice and dust respectively, and $X_{cr}$ the mass fraction of crystalline water ice. 
We also consider a Hertz factor $h$ with a fixed value of 0.1 to account for the granular structure of the solid \citep{Hue06}.
We finally use the Russel formula \citep{Rus35} to calculate a correction factor $\phi$ which should be applied to $\kappa _s$, to account for the effects of porosity \citep{Esp91, Cor97, Oro99}. It depends on the porosity $\psi$, and the ratio $f=\frac{\kappa_p}{\kappa_s}$, as
\begin{equation}
\phi = \frac{\psi ^{2/3} f +(1- \psi ^{2/3})}{\psi -\psi ^{2/3}+1-\psi ^{2/3}(\psi ^{1/3}-1)f},
\end{equation}
The material effective thermal conductivity is consequently:
\begin{equation}
\kappa = \phi ~ h ~ \kappa _s.
\end{equation}

\subsection{The object, the orbit and the albedo patch at the surface}
As mentioned in Section 2.1, the object is assumed to be a sphere, made of a porous matrix of amorphous water ice and dust. The thermal properties of such a mixture have been presented in the previous section. We also assume a radius $R$ of 2~km for the object, which is a typical comet radius \citep{AHe95}. 
Following the idea that there might be a dynamical cascade from the Kuiper Belt to the Centaurs to the JFCs, due to gravitational interactions with giant planets, we assume that the body enters the inner solar system on a Centaur-like orbit. 
The rotation period of the object is considered to be 10~hrs, typical of Centaurs and KBOs \citep{She08}. 

Finally, we assume that the initial object has a non-uniform Bond albedo at the surface. We consider a surface of 10\% Bond albedo ($\mathcal{A}_{surf}$), with a patch of 60\% Bond albedo ($\mathcal{A}_{patch}$), corresponding to a fresh ice/frost region on an otherwise dirty ice or refractory surface. While most TNOs and Centaurs have dark surfaces, some have high average albedos due to the presence of surface ice \citep{Sta08}. The patch is initially positioned between latitudes +22.5$^{\circ}$ and -22.5$^{\circ}$, and between longitudes 0$^{\circ}$ and 180$^{\circ}$. This corresponds to 1/8th of the overall surface, leading to a spherical average albedo of about 16\%. The parameters of the object, the orbit and the patch are summarized in Table \ref{obj}. For each simulation we also consider a reference case in which all the parameters are the same except that there is no albedo spot at the surface.

\section{Results}

In the interests of brevity, we focus the discussion on six cases as summarized in Table \ref{cas} (c.f. Figures \ref{radcr} to \ref{radex}). Cases A, D, E and F illustrate the influence of the heliocentric distance through variations of the semimajor axis and the eccentricity. Case B illustrates the effects of the obliquity, and Case C illustrates the influence of the material thermal conductivity. We performed additional simulations to explore the effects of the size and position of the patch or the albedo difference between the patch and the surface.

\subsection{Effects of non-uniform albedo}
\begin{figure}
\includegraphics[scale=0.9]{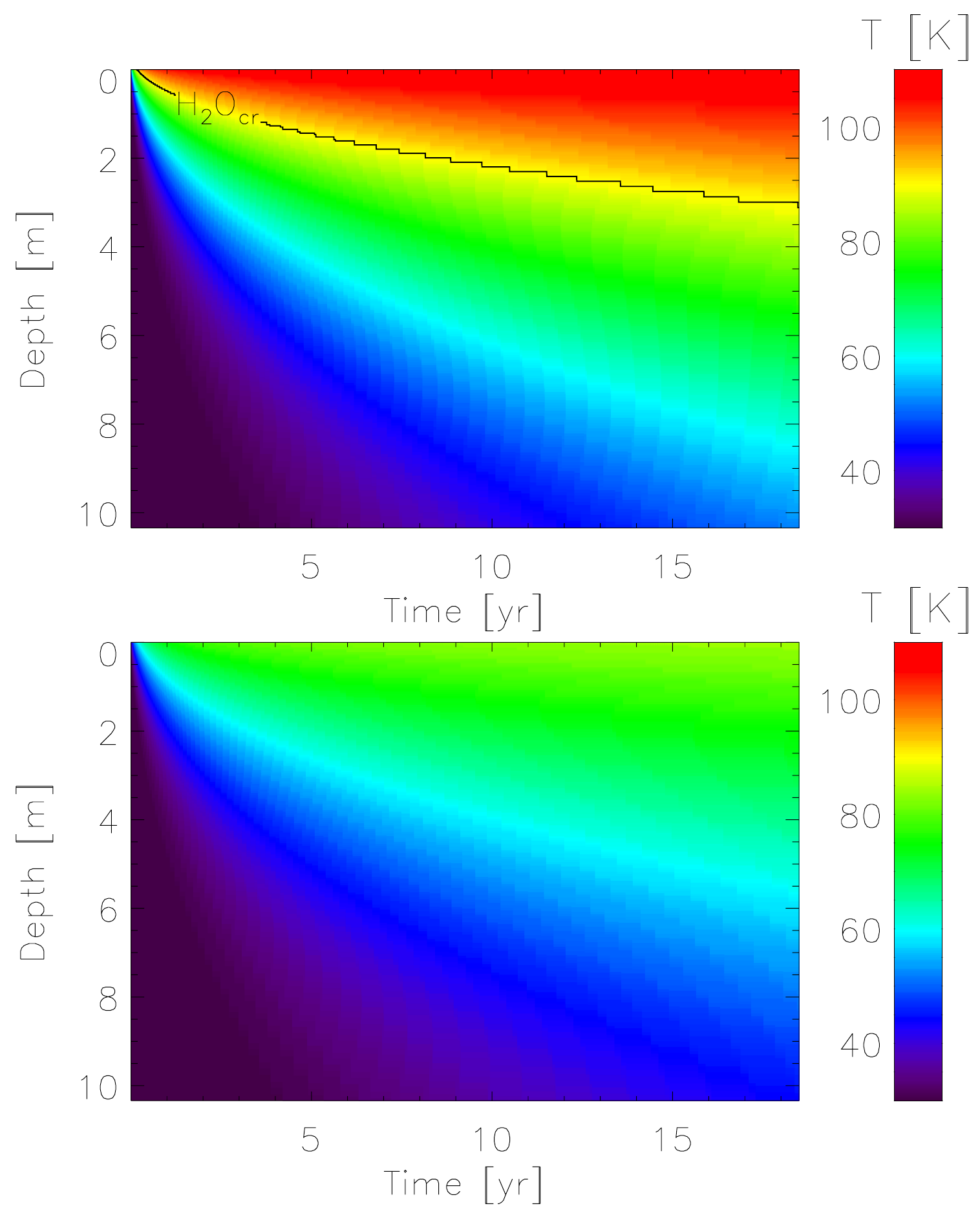}
\caption{\label{radcr}Case A. Radial evolution of the temperature under a given point on the equator: reference model (top panel) and inside the high albedo spot (bottom panel). The water ice crystallization boundary is delineated by the black line. No ice crystallizes under the spot.}
\end{figure}

\begin{figure}
\includegraphics[scale=0.9]{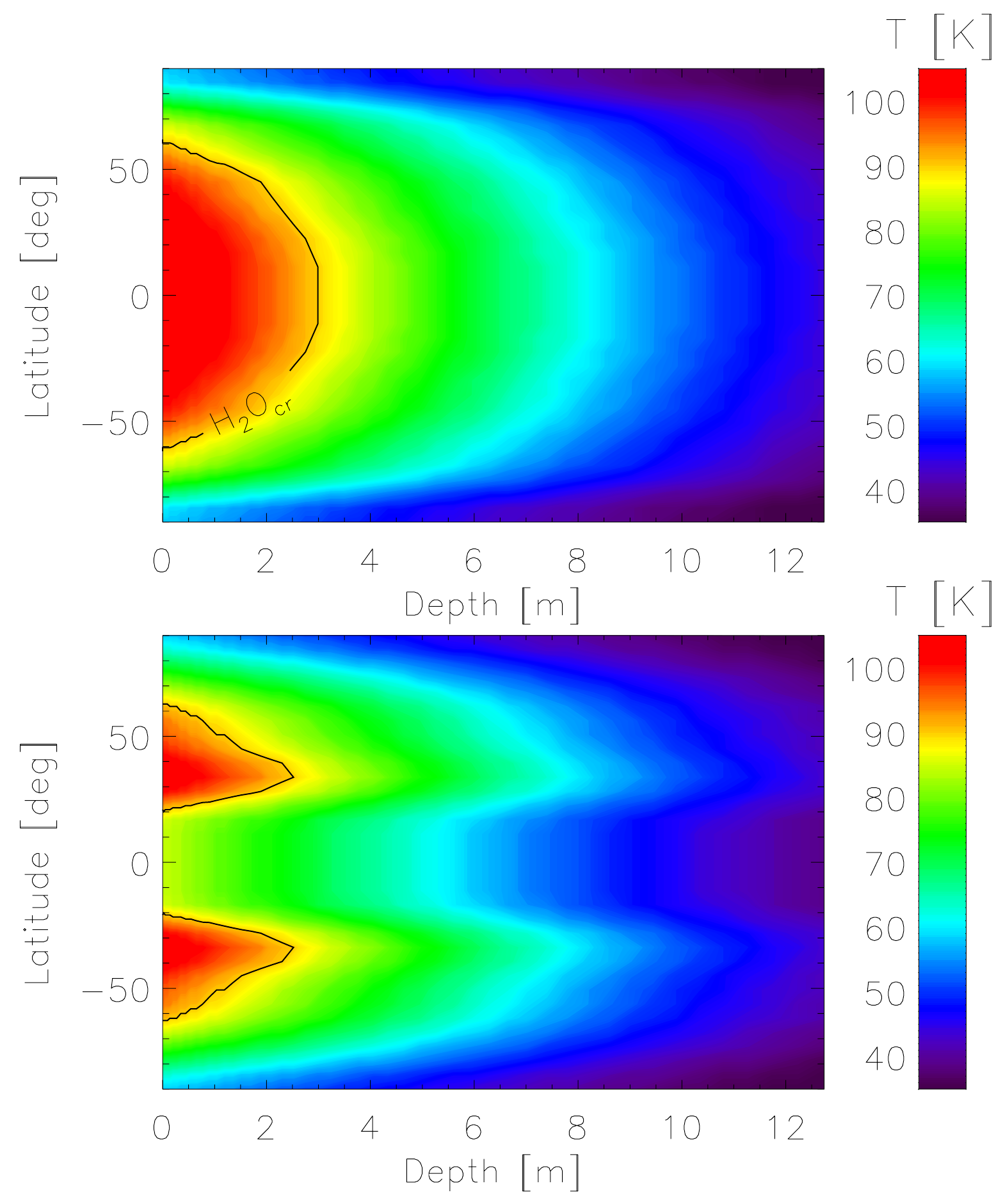}
\caption{\label{mercr}Case A. Temperature distributions along a meridian: reference model (top panel) and with a surface albedo spot located between latitudes +22.5$^{\circ}$ and -22.5$^{\circ}$ (bottom panel), after one orbital period. The water ice crystallization boundary is delineated by the black line.}
\end{figure}

\begin{figure}
\includegraphics[scale=0.9]{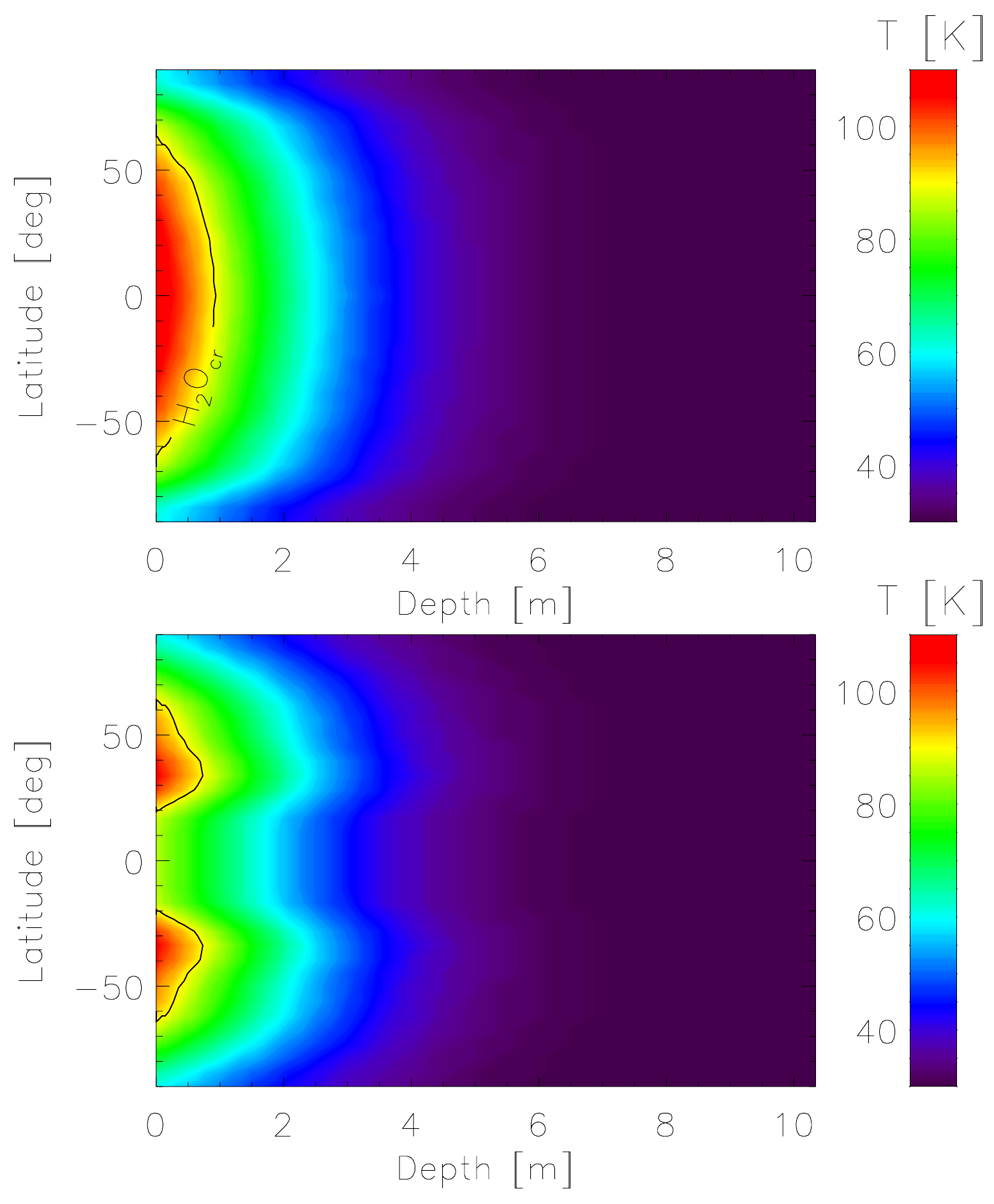}
\caption{\label{merklow}Case C. Same as Fig.\ref{mercr} with a thermal conductivity 10 times lower than in Case A, showing the reduced spatial scale of the thermal structure.}
\end{figure}

\begin{figure}
\includegraphics[scale=0.9]{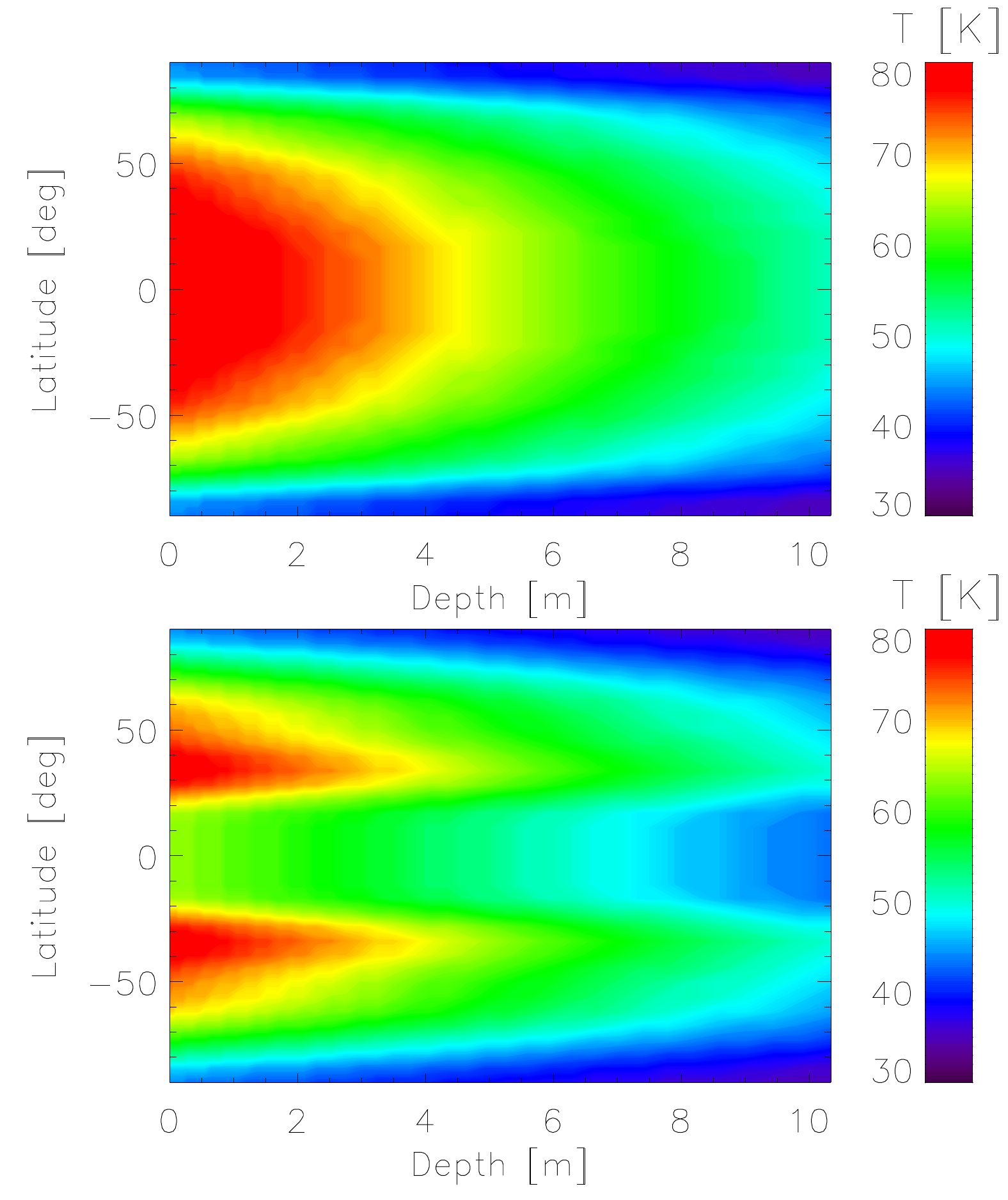}
\caption{\label{mer12}Case D. Same as Fig.\ref{mercr} with a=12~AU. The crystallization threshold is not reached at this larger distance.}
\end{figure}

\begin{figure}
\includegraphics[scale=0.9]{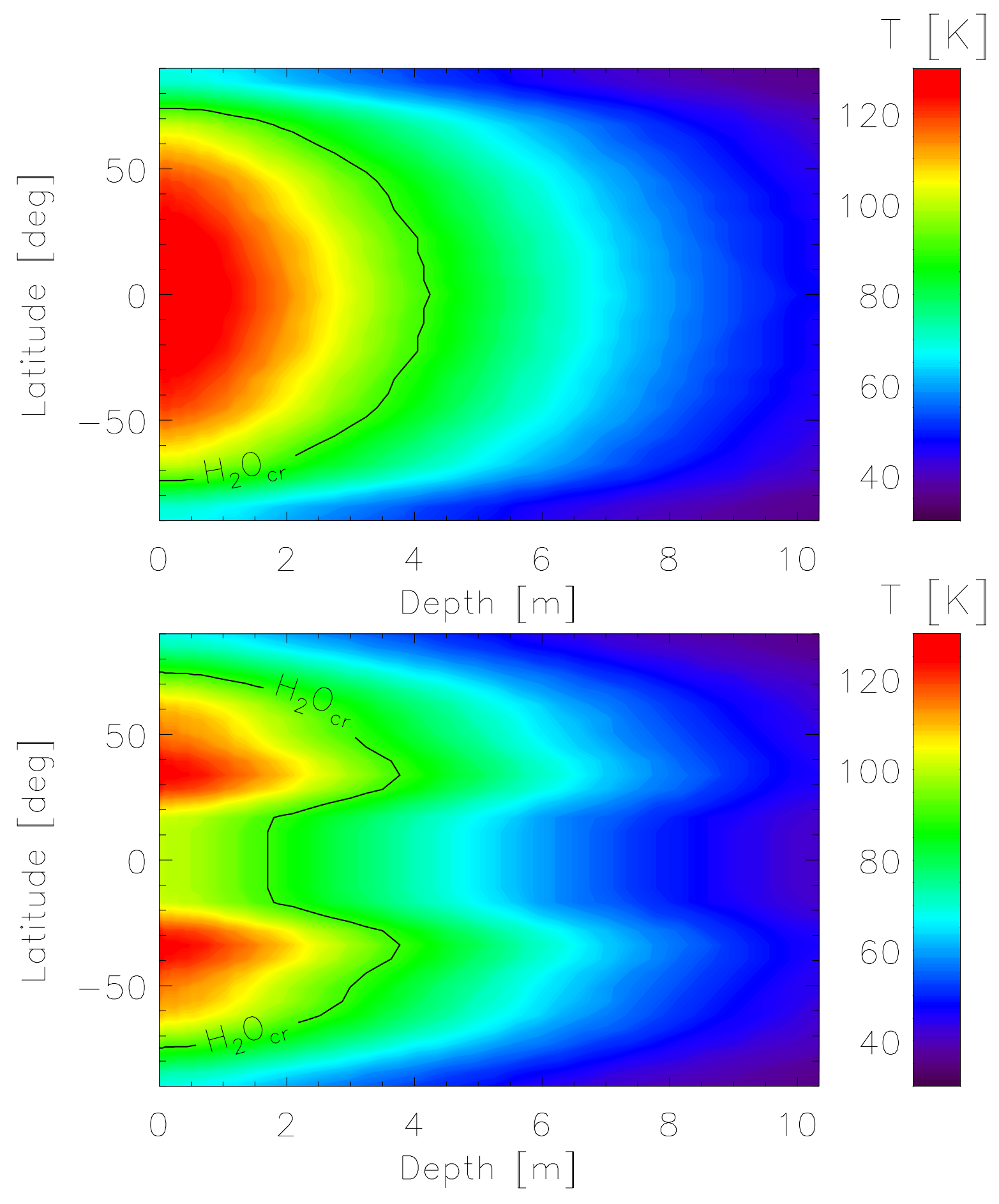}
\caption{\label{crist5}Case E. Same as Fig.\ref{mercr} with a=5~AU.}
\end{figure}

The spatially varying albedo induces a diurnally and annually modulated heat wave in the nucleus, with a peculiar shape that produces lateral subsurface thermal gradients. These effects are illustrated by Figs.\ref{radcr} and \ref{mercr} for Case A. Figure \ref{radcr} shows the evolution of the temperature beneath a given point on the equator with or without (reference case) the albedo spot. The temperature difference at the surface caused by the  spot is about 30~K. Figure \ref{mercr} shows that the region beneath the albedo patch remains 20 to 30~K cooler than in the reference case, creating a thermal shadow that appears very quickly when the object enters the inner solar system. On both figures, the black line labeled \textit{H$_2$O$_{cr}$ }stands as a limit beyond which amorphous water ice has been crystallized (in the hottest regions).

We expect that compositional gradients would also develop, following these lateral thermal gradients, because water ice crystallization and volatile sublimation are strongly temperature dependent. Specifically, cold regions in the thermal shadows of surface albedo features should be enhanced in volatiles relative to neighboring unshadowed regions, for two reasons.  First, uncrystallized ice in a local cold spot will retain its full complement of trapped volatiles, while these volatile species will have been liberated from surrounding crystallized ice.  Second, the thermal shadows may act as cold-traps in which volatile abundances are enhanced even further by the diffusive migration of molecules from crystallized ice adjacent to uncrystallized ice. 

We consider these points for Case A, in which crystallization is triggered by insolation both in the reference and non-uniform-albedo models, while the region located under the albedo spot remains cool enough to prevent crystallization (Fig.\ref{mercr}). 
Volatiles trapped in the amorphous matrix will be released upon crystallization \citep{Bar85, Lau87, Not96, Bar98, Not03} and travel in a free molecular flow (Knudsen flow, which is typical for comets, see \citet{Hue06}). The diffusion coefficient of gases released upon crystallization is given by \citet{Pri92}:
\begin{equation}
D=\frac{4}{3}vK_p,
\end{equation}
with $v=\sqrt{\frac{8 k_B T}{m \pi}}$ being the mean thermal gas velocity ($k_B$ is the Boltzmann constant, $T$ [K] the temperature and $m$ [g] the molecule mass), and $K_p$ a length coefficient which characterizes the porous material. For a medium made of randomly packed spheres with a resulting porosity $\psi$ and a pore radius $r_p$, $K_p=\frac{\psi^{3/2} r_p}{(1-\psi)^{1/3}}$ \citep{Pri92}. The pore radius is very uncertain. We here use r$_p$=10$^{-6}$~m but values an order of magnitude smaller are possible. Substituting $\psi$=30\% gives $K_p=1.85\times10^{-7}$~m. The diffusion coefficients of CO and CO$_2$ are $D_{CO}$=7.11$\times$10$^{-5}$~m$^2$~s$^{-1}$ and $D_{CO_2}$=5.67$\times$10$^{-5}$~m$^2$~s$^{-1}$ respectively, with T=110~K the maximum temperature reached within the crystalline regions. 

The gas diffusion length is given by $\ell=2\sqrt{Dt}$ with $t$ [s] the time. For example, in a typical nucleus rotation period of $\sim$10~hr, the molecules can flow through  $\ell \sim$3~m of the porous material given the above diffusion coefficients.   In one orbit (18.5 yr for a=7~AU), CO and CO$_2$ molecules can flow through $\ell \sim$400~m. Each thermal shadow is thus surrounded by a layer of radial extent $\ell$ from which liberated volatile molecules might be trapped.  The trapping will not be perfectly efficient, because liberated molecules can also migrate to the free surface of the nucleus and escape, or move down the thermal gradient into the cometary interior and re-freeze.  Nevertheless, we  expect that migration into thermal shadow cold-traps will produce a preferred spatial scale for volatile segregation, with the volatile enhancement being strongest for albedo spots having size comparable to the diffusion length.  For larger albedo spots, we expect that migrating volatiles will be trapped in a rim having thickness comparable to $\ell$.

On average, a Centaur can spend $\sim$10~Myr on its orbit before either leaving the solar system or becoming a JFC, due to gravitational interactions with giant planets \citep{Tis03, Hor04}. We performed simulations over 10~Myr, which showed that the subsurface cold plug can persist. 
After 10~Myr on the Centaur orbit, the thermal shadow produced by the presence of the higher albedo spot at the surface reaches about 400~m deep provided the nucleus spin vector remains constant over this period. Lateral heat fluxes only begin to erase the lateral thermal gradients, which are still important, in particular close to the surface where the temperature difference is about 20 to 30~K.

\subsection{Effect of obliquity}
\begin{figure}
\includegraphics[scale=0.9]{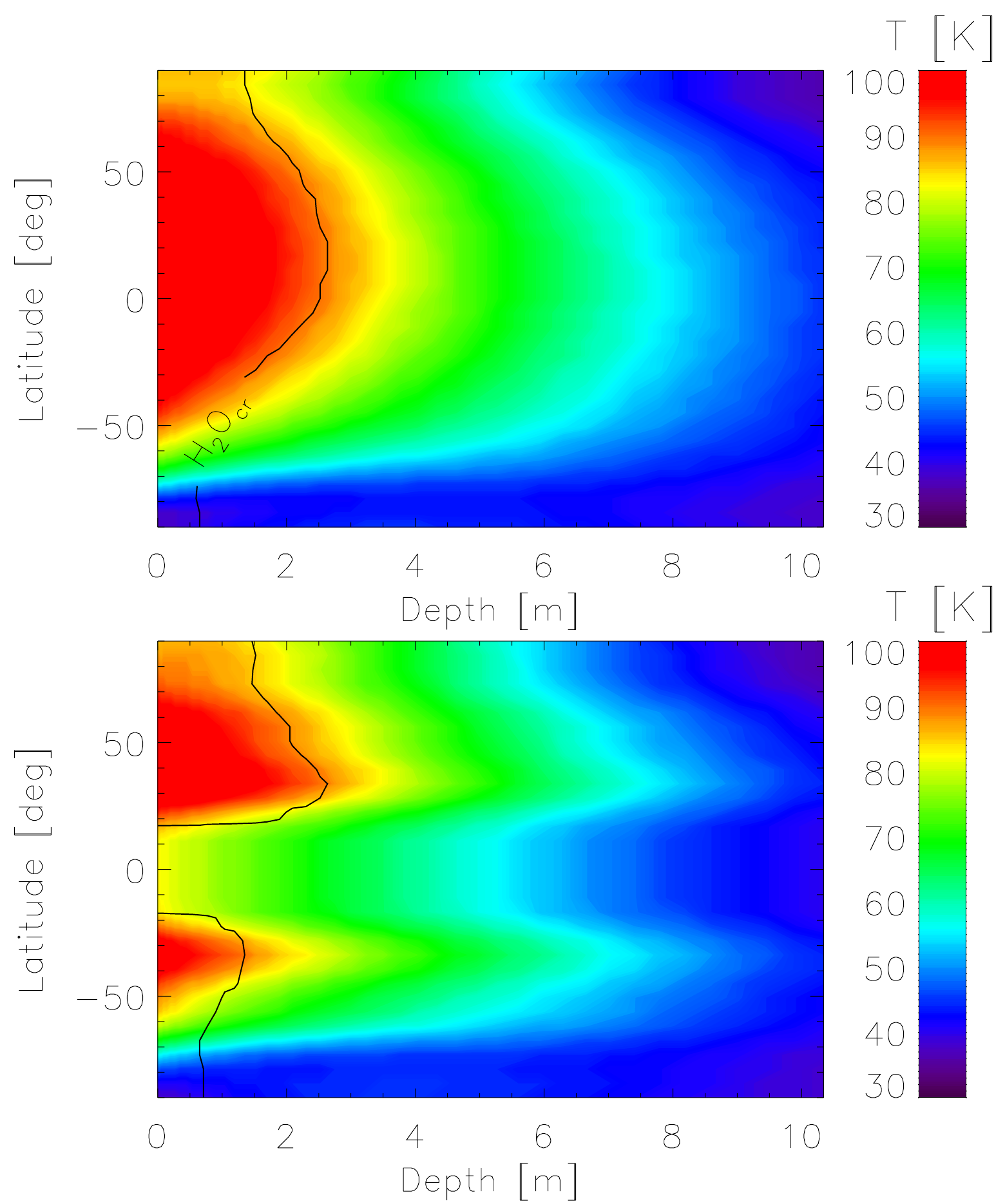}
\caption{\label{merobl}Case B. Same as Fig.\ref{mercr} showing asymmetries produced by obliquity 20$^\circ$. The subsolar point is located at the equator moving southward.}
\end{figure}

We found that obliquity is the parameter most affecting the formation of a thermal shadow.
At non-zero obliquity, the variations in the subsolar point latitude introduce asymmetry in the propagation of the heat wave, as illustrated by Case B in Fig.\ref{merobl} ($\Theta$ = 20$^{\circ}$). The temperature distribution with a non-zero obliquity varies across the orbit due to the variations of the subsolar point latitude, in addition to variations attributable to the propagation of the heat wave (which have been illustrated previously). Figure \ref{merobl} thus corresponds to a snapshot of the distribution, in which the subsolar point is located at the equator, and is moving southward.

After 10~Myr spent in the inner solar system, the temperature distribution for non-zero obliquity tends to become more uniform as the effects of the latitudinal movements of the subsolar point are averaged out. While the temperature difference between the poles and the equator is almost 40~K for $\Theta$ = 0$^{\circ}$, it is only 5~K with $\Theta$ = 45$^{\circ}$ after 10~Myr. 
Consequently, for high obliquities (typically larger than 45$^{\circ}$), the thermal shadow disappears with time, even if it appeared during the first few orbits. In the extreme case of $\Theta$ = 90$^{\circ}$, no thermal shadow ever appears despite the presence of temporary strong lateral thermal gradients. The radial propagation of the heat wave and lateral heat fluxes erase the potential compositional gradients already during the first orbit. 
Still we find that a subsurface thermal shadow can be maintained over 10~Myr on a Centaur-like orbit if the obliquity is low ($<30^{\circ}$ for substantial volatile enhancement effect). 

\subsection{Variations of other parameters}
The thermal conductivity affects mainly the radial extent of the cold plug, since it controls the efficiency of the heat transfer in the material, as illustrated by Case C and Fig.\ref{merklow}. In this case, we considered a thermal conductivity ten times lower than in the other cases. Interestingly, a low thermal conductivity implies that lateral heat fluxes are very ineffective in erasing any plug thus produced, which can survive for more than 10~Myr even if their lateral extent is initially small. 

\begin{figure}
\includegraphics[scale=0.9]{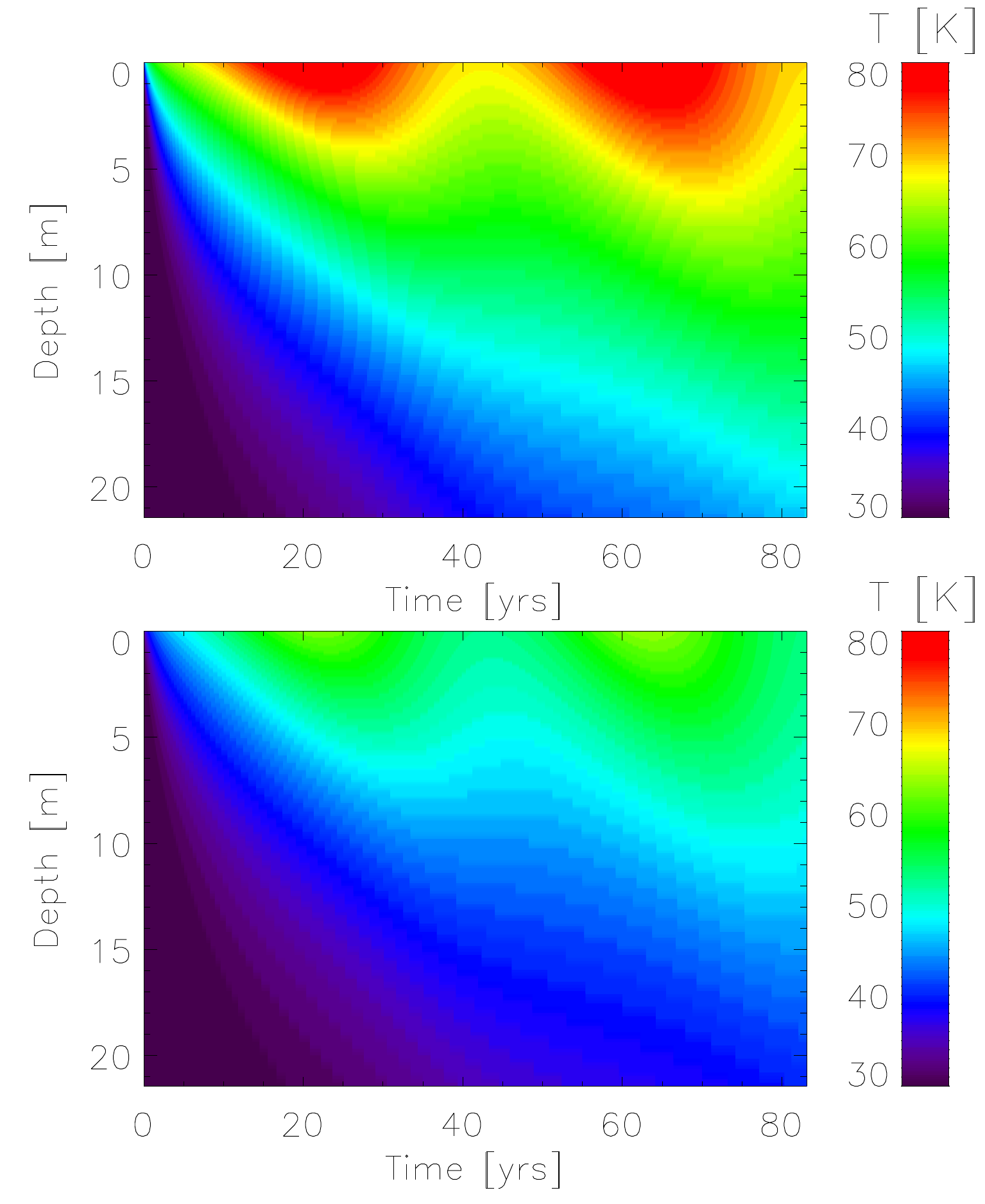}
\caption{\label{radex} Case F. Same as  Fig.\ref{radcr} for an eccentric orbit (a=15~AU, e=0.2).}
\end{figure}

The heliocentric distance is also an important factor. With increasing heliocentric distance, the steepness of the lateral thermal gradients decreases. In addition, as shown by Fig.\ref{mer12} for Case D, crystallization might not be reached in the hottest regions, thus limiting the volatile enhancement in the thermal shadows of surface features. On the contrary, a smaller heliocentric distance will induce higher surface temperatures. In Case E, the crystallization threshold is reached also under the albedo spot (Fig.\ref{crist5}). Over the time of residence in the giant planets region, the non-uniform structure could still exist. The most volatile enriched region would nonetheless be located a few meters to a few tens of meters deep. We found that the orbital eccentricity has very limited effect (Case F). Its influence is restricted to the amount of energy to be transferred to the subsurface (see Fig.\ref{radex}), which varies around the orbit. This merely impacts the shape or extent of the temperature distribution in the case we show in Fig.\ref{radex}. The orbit that we considered has a semi-major axis of 15~AU and an eccentricity of 0.2, resulting in a perihelion distance of 12~AU, the same distance as in Case D for a circular orbit with $a$=12~AU. Larger eccentricities could nonetheless have more influence, as the energy provided close to perihelion could potentially trigger the crystallization of amorphous water ice. 

The surface of a comet is impacted by a large variety of energetic particles, which might produce an irradiation crust \citep{Str91, Hud08}. The first few centimeters are the most affected, but high energy particles might penetrate up to a meter beneath the surface. The thermal properties of such a crust are not constrained yet, but recent laboratory experiments on porous dust aggregates indicate thermal conductivity between 10$^{-3}$ and 10$^{-2}$~Wm$^{-1}$K$^{-1}$ \citep{Kra11}. This is very  similar to the thermal conductivity of the material we are considering. Therefore, this crust would not prevent the progression of the heat wave toward the center of the object. It would have a damping effect, moving the interesting boundary a meter deeper, and inducing lower temperatures. The extent of non-uniform structures could be limited in this case, but would still exist, especially for objects orbiting close to the Sun (as in Case E).

\section{Discussion}
Our simulations show that non-uniform surface albedo creates thermal gradients in the subsurface layers, which can produce a long-lived non-uniform subsurface structure.
The emergence of  compositional non-uniformity depends on the albedo difference between the surface and the patch, rather than the albedo itself. Local albedo variations of a factor of four exist on comet 103P/Hartley~2 \citep{AHe11}, suggesting that thermal shadow effects could be very strong. Although we considered a large albedo spot, smaller scale surface features would have the same impact on the subsurface, since with low thermal conductivities, lateral heat transfers are quite inefficient in erasing the cold plugs. The overall albedo could remain very low, as the variation attributable to the spot would be limited. 
Real comet nuclei have complex shapes and surface features, as revealed by spacecraft observations like for comets 9P/Tempel~1 or 103P/Hartley~2 (Deep Impact and EPOXI missions respectively). The effects of surface topographic features like craters would mimic those of albedo features, since ultimately the important factor is the heat transferred to the subsurface. Consequently, the presence of craters, topographic features, boulders or any other source of shadowing can generate subsurface lateral thermal gradients and non-uniform compositions. 

If the crystallization threshold is locally reached, the thermal shadow could become strongly enriched in volatiles, while the surrounding crystallized ice would be depleted in volatiles. In our cases, the super-volatile CO would most likely escape the body, while a less volatile compound such as CO$_2$ could recondense in the thermal shadow. The diffusion length is independent of the patch size, and depends mainly on the material thermal and structural properties. Time is also an important factor. If the size of the spot is comparable to or smaller than the diffusion length, the volatile abundance enhancement of the cold plug could be extremely important. Such features have been reported to have a scale of tens to hundreds of meters \citep{Mum93, Wei04, AHe11}, which is very similar to the diffusion lengths considered here.
Consequently, we can predict that the scale of the enriched regions would range from a few meters to a few hundred meters, depending on i) the size of the patch, as the size of the enriched region cannot be larger than the cold plug, ii) the material properties and iii) the time the molecules had to flow in the porous medium. Nonetheless, the gas phase is not accounted for nor modeled in our simulations. The effects of such a gas phase could strongly modify our results, as it can locally affect the thermal conductivity, the porosity or pore sizes. In addition, instabilities caused by pressure build-up could develop and blow up some surface layers. There are, however, too many unknowns to meaningfully model all these processes.

The compositional variations produced in the subsurface can be sustained until the body becomes a JFC. In this case, the resulting internal composition would be strongly non-uniform, and the cometary activity would be generated through jets. These would be produced as the object orbits closer to the Sun, where insolation can finally trigger the sublimation of volatiles which were concentrated in the thermal shadows, and/or the crystallization of these regions. Consequently, we expect that non-uniform thermal and compositional structure should be common. Identification of compositional differences in a single nucleus does not necessarily imply that the comets were built from cometesimals formed at different heliocentric distances, with distinct compositions. 

\section{Summary}

Fast, 3-dimensional thermal evolution simulations show:

\begin{itemize}
\item[1.] Non-uniform surface albedos on comets can generate long-lived thermal shadows in the immediate sub-surface regions. 

\item[2.] Temperature-sensitive processes (including sublimation and crystallization) proceed at different rates inside and outside the thermal shadows, leading to the development of volatile-enhanced shadow cold-traps.

\item[3.] Compositional gradients caused by thermal shadows should be most pronounced for albedo spot sizes  comparable to the diffusion length (typically from a few meters to a few hundred meters for structures growing on the rotational and orbital timescales). 

\item[4.] Under some circumstances, subsurface temperature structure can be preserved for the mean lifetime of a Centaur (10~Myr) before the object becomes a Jupiter Family Comet.

\item[5.] Observations of jets and non-uniform compositions in cometary nuclei do not necessarily imply an initially non-uniform composition.

\end{itemize}

\acknowledgments
We thank O. Groussin and H. Hsieh for valuable comments on the manuscript.
This research was supported by a NASA Herschel grant to David Jewitt. 

\clearpage

\begin{table}[ht!]
\begin{center}
\caption{\label{param}Heat capacities and thermal conductivities of different components in the material mixture.}
\begin{tabular}{clcl}
\hline
Param. & Value & Unit & Ref. \\
\hline \hline
$c_{H_2O}$     & $7.49T+90$                                 & Jkg$^{-1}$K$^{-1}$ & G\&S36\\
$c_{d}$             & $1200$                                         & Jkg$^{-1}$K$^{-1} $& E\&S83\\
$\kappa _{a}$  & $2.34~10^{-3}T+2.8~10^{-2}$ & Wm$^{-1}$K$^{-1}$ & Kl80\\
$\kappa _{cr}$ & $567~/~T$                                   & Wm$^{-1}$K$^{-1}$ & Kl80\\
$\kappa _{d}$  & $4.2$                                             & Wm$^{-1}$K$^{-1}$ & E\&S83\\
\hline
$c_{init}$             &$760$                                          &Jkg$^{-1}$K$^{-1} $&\\
$\kappa _{init}$  &$6.17~10^{-2}$                          &Wm$^{-1}$K$^{-1}$&\\
\hline
\end{tabular}
\end{center}
\tablecomments{Heat capacities: $c_{H_2O}$ for water ice and $c_{d}$ for dust respectively, thermal conductivities: $\kappa _{a}$ and $\kappa _{cr}$ for amorphous and crystalline water ice respectively, and $\kappa _{d}$ for dust. $c_{init}$ and $\kappa _{init}$ correspond to the initial values of the heat capacity and thermal conductivity in the simulations.}
\tablerefs{G\&S36: \citet{Gia36}, E\&S83: \citet{Ell83}, Kl80: \citet{Kli80}. } 
\end{table}

\clearpage

\begin{table}[ht!]
\begin{center}
\caption{\label{obj} Initial values for various parameters relative to the object, its orbit and the non-uniform surface albedo, considering Case A.}
\begin{tabular}{l|cccccccccc}
\hline
Param. & R & $\rho_{bulk}$ & $\mathcal{A}_{surf}$ & $\mathcal{A}_{patch}$&$\alpha^{\star}$ & a & e & $\Theta$ & P$_{rot}$ & P$_{orb}$\\
\hline \hline
Unit       & km & g cm$^{-3}$ &-&-&-&AU&-&${}^{\circ}$&hrs&yrs\\ 
Value   & 2&1&10\%&60\%&12.5\%&7&0&0&10&18.5\\
\hline
\end{tabular}
\end{center}
\tablecomments{$\star$: area ratio between the spot and the overall surface.}
\end{table}

\clearpage

\begin{table}[ht!]
\begin{center}
\caption{\label{cas} Parameters for the different cases illustrated, with the corresponding figures.}
\begin{tabular}{ccccccl}
\hline
Case & $\kappa_{init}$ & a  & e & $\Theta$  & Fig. & Description\\
          &  [Wm$^{-1}$K$^{-1}$]  & [AU] & & [${}^{\circ}$] & \\
\hline \hline
A & 6.17~10$^{-2}$ & 7   & 0     & 0    & \ref{radcr} &T radial evolution with time\\
   &                              &       &        &       & \ref{mercr} &T distrib. after one orbital per.\\
B & 6.17~10$^{-2}$ &   7 & 0     & 20  &\ref{merobl}  &T distrib. after one orbital per. \\
C & 6.17~10$^{-3}$ &  7  & 0     &   0  & \ref{merklow} &T distrib. after one orbital per. \\
D & 6.17~10$^{-2}$ & 12 & 0     &  0   &\ref{mer12}  &T distrib. after one orbital per. \\
E & 6.17~10$^{-2}$ &   5 & 0     &  0   &\ref{crist5}  &T distrib. after one orbital per. \\
F & 6.17~10$^{-2}$ & 15 & 0.2 & 0     & \ref{radex} &T radial evolution with time\\

\hline
\end{tabular}
\end{center}
\end{table}

\clearpage

\clearpage

\end{document}